\documentclass[12pt]{article}
\usepackage{amsmath,amssymb,graphicx} 
\usepackage{epsf}
\usepackage{subfigure}
\usepackage{cite}

\def\beq{\begin{equation}}
\def\eeq{\end{equation}}
\def\bea{\begin{eqnarray}}
\def\eea{\end{eqnarray}}
\def\beal{\begin{align}}
\def\eeal{\end{align}}

\def\as{\alpha_s}
\def\ren{{\rm ren}}

\def\vep{\varepsilon}
\def\W{{\cal W}}
\def\F{{\cal I}}

\begin{document}
\begin{titlepage}

\begin{flushright}
YITP-SB-10-26\\
\end{flushright}

\vskip.5cm
\begin{center}
{\large \bf  Diagrammatic Exponentiation for Products of Wilson Lines}
\vskip.2cm
\end{center}

\begin{center}
  {Alexander Mitov}, {George Sterman}, {Ilmo Sung} \\

\vskip 8pt

 {\it \small C.N. Yang Institute for Theoretical Physics\\
Stony Brook University, Stony Brook, New York 11794-3840, USA}\\

\vspace*{0.853cm}

{\bf Abstract \vspace*{-0.1cm}\\}\end{center}

We provide a recursive diagrammatic prescription for the
exponentiation of gauge theory amplitudes  involving products
of Wilson lines and loops.
This construction generalizes the concept of webs, originally
developed for eikonal form factors and cross sections with
two eikonal lines,
to general soft functions in QCD and related gauge theories.
Our coordinate space arguments 
apply to arbitrary  paths for the lines.

\end{titlepage}
\newpage

\section{Introduction}
\label{intro} \setcounter{equation}{0} \setcounter{footnote}{3}

Path-ordered exponentials of gauge fields over space-time curves are an essential
ingredient in many descriptions of gauge theory dynamics.  A generic form
for these functions is \cite{Bialynicki-Birula,Wilson:1974sk}
\bea
\Phi^{(f)}_C(y,x)
=
P\, \exp\, \left[\, \int_0^\infty d\tau\, \xi_C(\tau)\cdot A^{(f)}\left(\xi_C(\tau)\right)\, \right]\, ,
\label{oe}
\eea
where as $\tau$ varies, the coordinate $\xi^\mu_C(\tau)$ traces curve $C$
from point $x=\xi(0)$ to point $y=\xi(\infty)$, which may be
open or closed ($y=x$).    The gauge field may be in any
representation $f$ of the group.   Such ordered exponentials are generally
referred to as Wilson loops and lines.
Matrix elements of products of Wilson lines have
become a familiar feature of factorized amplitudes \cite{Kidonakis:1998nf,Sen:1982bt} and
resummed cross sections in QCD \cite{Kidonakis:1997gm}  and
related gauge theories \cite{Alday:2008yw}.\footnote{In particular, ``soft functions", which organize non-collinear
soft radiation, are also a key part  of factorizations \cite{Lee:2006nr} and
resummations based on direct analyses in perturbative QCD
and soft-collinear effective theory \cite{Fleming:2007xt,Chiu:2009mg}.}
Applications to physical process with an electroweak hard scattering
are generally based on products of two Wilson lines that meet at a point
in a color singlet configuration.   This vertex defines a composite
operator that requires renormalization, and is often termed the ``cusp" vertex \cite{Polyakov:1980ca}.   
The description of QCD hard scatterings \cite{Kidonakis:1998nf,Sen:1982bt}
generally requires several lines  to meet at
what we will call a ``multi-eikonal vertex'', of which the
simplest example is the cusp.\footnote{Generally, the term `eikonal' refers to
a source with constant velocity.  From the point of view of renormalization,
however, it is only necessary that the lines meeting at a point have
a well-defined local velocity, or tangent vector.   We will therefore
use this terminology even though our considerations below apply to lines
whose local velocities are otherwise arbitrary.}    
Providing our Wilson lines with color indices, a four-line multi-eikonal
vertex, for example, can be represented by a constant matrix, $c_I$ in color space
that links the indices \cite{Kidonakis:1998nf,Sen:1982bt},
\bea
W^{[\rm f]}_I{}_{\{r_k\}}
&=&
\sum_{\{d_i\}}
\langle0|\, \Phi_{v_4}^{[f_4]}(\infty,0)_{r_4,d_4}\; 
\Phi_{v_3}^{[f_3]}(\infty,0)_{r_3,d_3}\cr
&\ & \hspace{15mm} \times
\left( c_I\right)_{d_4d_3,d_2d_1}\; 
\Phi_{v_1}^{[f_1]}(0,-\infty)_{d_1,r_1}
\Phi_{v_2}^{[f_2]}(0,-\infty)_{d_2,r_2}\, |0\rangle \,.~~~~~
\label{eq:wivertex}
\eea
For the eikonal Wilson lines of this expression, constant velocities $v_i$ label the curves,
which we can choose to be $\xi_j(\tau_j) = v_j\tau_j$ (outgoing) or $\xi_i(\tau_i)=-v_i/\tau_i$ (incoming).   The curves meet at the origin,
either from infinity in the past or to infinity in the future.

Amplitudes of multi-eikonal vertices 
contain the bulk of information necessary to reconstruct 
the full infrared structure of multiparton amplitudes in QCD \cite{Sterman:2002qn}.
An even more direct connection can be found in ${\cal N}=4$ 
supersymmetric Yang-Mills theory (SYM), for which certain closed loops of Wilson 
lines with sequential cusp vertices have the remarkable property of being dual
descriptions to full partonic scattering amplitudes \cite{Drummond:2007aua},
a feature with suggestive connections to the strong-coupling
limit \cite{Alday:2007hr}.

Many of the useful features of the examples mentioned above
result from their exponentiation properties \cite{Sterman:2002qn,Magnea:1990zb}.
These may be formulated quite generally in terms
of the anomalous dimensions of multi-eikonal vertices \cite{Kidonakis:1998nf,Polyakov:1980ca}.
For the cusp anomalous dimension, however, there is a
more detailed exponential form, in which 
the singlet product of Wilson lines is written as the
exponential of a sum of 
two-eikonal irreducible
diagrams with modified
color factors, the so-called ``webs" \cite{Gatheral:1983cz}.    The renormalization
of the cusp anomalous dimension can be reformulated
simply in terms of the webs, and many regularities
at nonleading order flow from the underlying exponentiation.
The original proofs \cite{Gatheral:1983cz} of diagrammatic exponentiation were formulated
in momentum space, specifically for the cusp vertex and for
Wilson lines of fixed velocity.   These considerations
were revisited and extended recently in Ref.\ \cite{Laenen:2008gt}.
In the present paper, we will develop a diagrammatic exponentiation,
not only for multi-eikonal scattering, but also 
for arbitrary products of Wilson lines of arbitrary
length, which may or may not meet at cusp
or multi-eikonal vertices.   

A potential application of our results is to the systematic
calculation of anomalous dimensions for 
multi-eikonal vertices.     The cusp anomalous dimension, for example,
can be determined directly from the webs, which
can streamline calculations \cite{Berger:2003zh,DelDuca:2010zp}.   
More general composite vertices, involving multiple Wilson lines 
require matrices of anomalous dimensions.   By now, these
products are understood to two loops for massless \cite{Aybat:2006mz}
as well as massive lines \cite{Kidonakis:2009ev,Ferroglia:2009ep,Mitov:2010xw}, and interesting progress has been made at higher orders of the massless case \cite{Becher:2009qa,Gardi:2009qi,Dixon:2009ur}.
In what follows, we will show that as for the cusp,
diagrammatic exponentiation can simplify, and certainly clarifies,
the calculation of anomalous dimensions beyond one loop.
As in Ref.\ \cite{Mitov:2010xw}, we will find it useful to
work in a coordinate space representation.

Graphical exponentiation also has potential applications
in the phenomenological treatment of cross sections.
Because the underlying structure is 
fundamentally nonperturbative, webs have been used to organize
the structure of power corrections due to 
soft radiation for Drell-Yan and related cross sections \cite{Laenen:2000ij}.
We anticipate analogous applications to QCD hard
scattering processes.

We present our diagrammatic construction in Sec.\ \ref{sec:unren}, with a
discussion of general diagrams for vacuum 
expectation values of products of Wilson lines.
We give a simple combinatorial identity for
products of Wilson line amplitudes considered 
directly in the coordinate space that contains their
defining curves.    From this identity, we derive an
iterative construction for the logarithm of the 
amplitude, based purely on considerations of counting.
We lose the simple condition of diagrammatic
irreducibility characteristic of  webs for the cusp, 
but in the general case the color factors of the subdiagrams that make up the logarithm are
intertwined.  As a result, we will
continue to use the term ``web" below to describe the 
result of this procedure.
The mixing of color structure gives
the general product a more complex form than 
the familiar case based on the cusp vertex.   
In this special case, however, the classic web formula 
is readily derived.  
In Sec.\  \ref{sec:ren}, we analyze the renormalization of multi-eikonal vertices
as they occur in the diagrams discussed in Sec.\ \ref{sec:unren},
and describe some features of renormalized webs for multiple lines.
We go on in Sec.\ \ref{sec:extend} to conclude with a brief discussion the extension of the formalism 
to cross sections and to the massless case.

\section{Coordinate Identities and Diagrammatic Exponentiation}
\label{sec:unren}

\begin{figure}
{\hskip 1.75 in \epsfxsize=8 cm \epsffile{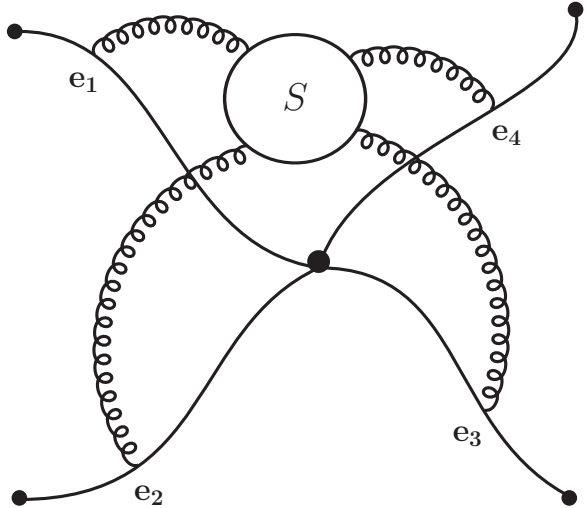}}\\
 \caption{Representation of the eikonal amplitude $A$ discussed
 in the text.   Each gluon line represents an arbitrary number of
 connections to each Wilson line, indicated by $e_a$ for line $a$, with 
 $a=1\dots L$ for $L$ Wilson lines.}  
 \label{fig:illustrate}
\end{figure}

We will be interested in the sums of diagrams of the sort illustrated in Fig.\ \ref{fig:illustrate},
in which an arbitrary number of Wilson lines (four in the figure) are connected
by gluon attachments in all possible ways.   The attached gluons may interact
in an arbitrary fashion, as indicated by the bubble $S$ in the figure.  For the purposes
of this argument, we denote the sum of all such diagrams for a set of 
Wilson lines over smooth paths $C_i$, $i=1\dots L$, as 
\bea
A[C_i]= \sum_{N\ge 0} A^{(N)}[C_i]\, ,
\eea
where the superscript $N$ denotes the order in $\alpha_s$.   For convenience, we absorb 
$(\as/\pi)^N$ into $A^{(N)}$ rather than to show it explicitly.   We also
suppress color indices, but generally the amplitude in Eq.\ (\ref{eq:wivertex}) is a vector in the space of color
tensors \cite{Kidonakis:1998nf}.   We will assume that
the curves $C_i$ may meet at one or more multi-eikonal vertices, although they need not do so.
We shall assume that they are otherwise nonintersecting.   The curves need not
be of infinite extent, although they may be.

In this section, we will treat the Wilson lines as they appear in 
perturbation theory, including in principle renormalization of all
terms in the Lagrange density, but not for the composite multi-eikonal
vertices.   Thus, we assume that the diagrams are regularized for
both ultraviolet and infrared divergences, and treat
them as convergent integrals.   We return to 
the renormalization of the composite vertices in the next section.

Of course, we can always write the amplitude formally as an exponential,
\bea
A[C_i]= \exp\, \left( \, w[C_i]\, \right)\, , \quad w[C_i] = \sum_{i\ge 1}w^{(i)}[C_i]\, ,
\eea
for some matrix $w[C_i]$, the logarithm of $A[C_i]$, which can also
be expanded in powers of $\as$, beginning at order $\as$ as indicated.  Both $w$ and $A$ are functionals of
the curves $C_i$, as shown, but we shall suppress this dependence as well below.

Our goal is to characterize the matrix $w$
order-by-order in perturbation theory for a general set of Wilson lines (that is, curves $C_i$).  
We will refer to diagrams that contribute to $w^{(i)}$ 
as ``webs" by analogy to the form factor \cite{Gatheral:1983cz}, although the diagrammatic
structure that will emerge from the reasoning below generalizes
the case of the cusp vertex.

We actually know the webs at lowest order, $w^{(1)}$,  
clearly given by the sum of all possible single-gluon exchanges
between the curves $C_i$, including the case of self-energies. 
It is therefore natural to pose our construction in recursive terms.
 
Suppose then, that we know the $w^{(i)}$ up
to some fixed order $N$.\footnote{We should point out that our construction could be
organized by treating  $N$ as simply the total number of vertices
at which gluons attach to Wilson lines.  Such an alternative treatment
would emphasize the arbitrary interactions between soft gluons
indicated in Fig.\ \ref{fig:illustrate}.  Indeed, these interactions
need not even be purely perturbative, or restricted to the choice
of four dimensions.   We choose not to follow this route because
the total loop order has a more direct relation to perturbative renormalization
for the multi-eikonal vertices.}   This knowledge is enough to
construct $A$ up to $N$-loop order.   
That is, we can write
\bea
\sum_{j=1}^N A^{(j)} = \exp \left(\sum_{i=1}^N w^{(i)} \right) + {\cal O}(\as^{N+1})\, ,
\label{eq:AN}
\eea
with corrections at the next order.
To relate the $N$th order expression in
  Eq.\ (\ref{eq:AN}) to $w^{(N+1)}$, we next expand the exponential, 
  remembering that the $w$'s are matrices,
\bea
\exp \left(\sum_{i=1}^N w^{(i)} \right)
&=&
\sum_{m=1}^\infty \frac{1}{m!}\, \left(\, \sum_{i=1}^N w^{(i)}\, \right)^m
\nonumber\\
&=& 
\sum_{m=1}^\infty \frac{1}{m!}\,\,
\sum_{i_m=1}^N\dots \sum_{i_1=1}^N\,
w^{(i_m)}\, w^{(i_{m-1})} \dots w^{(i_1)}\, .
\label{eq:productform}
\eea
Of course, some of these $w$'s may be
identical.   We will denote the number of copies of $w^{(i_c)}$ as  $m_c$
where $m_c$ can take any integer value, including zero.

Now consider the $N+1$st order, which we write in two ways.  The first
is as the sum of all $N+1$st order diagrams, denoted by $D^{(N+1)}$,
\bea
A^{(N+1)} &=& \sum_{D^{(N+1)}}  D^{(N+1)} \, .
\label{AasD}
\eea
The second expression for $A^{(N+1)}$ is as the 
$N+1$st order term from the exponential of $W$ up to order $N+1$,
\bea
A^{(N+1)} &=& \left(\, \exp\left[\sum_{i=1}^{N+1} w^{(i)} \right] \, \right)^{(N+1)}\, .
\eea
We can now solve for $w^{(N+1)}$ in terms of the sum of $N+1$-order diagrams;
it is only necessary to subtract from the sum over $D^{(N+1)}$ the
expansion of the exponential of the sum of $w^{(m)}$s up to $m=N+1$.  Using
Eq.\ (\ref{eq:productform}) for the expansion of the exponential,
and observing that $w^{(N+1)}$ can appear only by itself in the sum, we find
\bea
w^{(N+1)} = 
\sum_{D^{(N+1)}}  D^{(N+1)} - \left[\, \sum_{m=2}^{N+1} \frac{1}{m!}\,\,
\sum_{i_m=1}^N \dots \sum_{i_1=1}^N\,
w^{(i_m)}\, w^{(i_{m-1})} \dots w^{(i_1)} \right]^{(N+1)}\, .
\label{eq:first_wN+1}
\eea
Thus, the $N+1$st order contribution to the exponent is just
what is left over from the sum of all diagrams at that order when we
subtract the $N+1$st order result of the exponentiation of lower orders.
In the following, we will learn how to
interpret these products of the $w$'s, which will allow us to write
a form that is more informative than Eq.\ (\ref{eq:first_wN+1}).

The $w^{(i)}$ are sums of diagrams with different attachments 
of gluons to each of the Wilson lines, and we will find it convenient
to group together those diagrams at each order with definite numbers of gluons, 
$e_a$, attached to the $a$th Wilson line, $a=1\dots L$.
Clearly, the values of $e_a$ are limited by the total loop order,
\bea
2\le \sum_{a=1}^L e_a \le 2N\, ,
\label{ealimit}
\eea
at $N$th order.
To label the possible connections at each order explicitly, we write
\bea
w^{(i)} = \sum_{E}
w^{(i)}_E\, ,
\label{eq:sumE}
\eea
where each $E=\{e_1\dots e_L\}$ is a member of the set of possible
assignments of the $e_a$, subject to (\ref{ealimit}).
In fact, the information contained in the subscript $E$ is all
we will need to know about the webs.   To see this, we consider
their general form as integrals over Wilson lines.

Each term $w^{(i)}_E$ is itself an integral over 
the positions of each of its external gluons along 
the Wilson line to which it attaches.   By construction, $w^{(i)}_E$ includes
a sum over all 
web
diagrams with $e_a$ gluons attached
to the $a$th line, so that  by Bose symmetry it is symmetric under
permutations of the gluons attached to each line.    Thus, without
loss of generality we order the parameters $\tau_j^{(a)}$, $j=1\dots e_a$
for each of the Wilson lines, 
defined as in Eq.\ (\ref{oe}),
and write
\bea
w^{(i)}_E
&=&
\prod_{a=1}^L\, 
\prod_{j=1}^{e_a}\int_{\tau^{(a)}_{j-1}}^\infty d\tau^{(a)}_j\,
\W_E^{(i)}\left(\{\tau^{(a)}_j\}\right)
\nonumber\\
&\equiv& \F_E\, [\W_E^{(i)}]\, ,
\label{eq:calFcalW}
\eea
where in the first equality we define $\tau^{(a)}_0\equiv0$.
The second form represents  the
integrals as a functional
$\F_E$, acting on  the ``internal web function" $\W_E^{(i)}$ corresponding
to $w^{(i)}_E$.  $\W_E^{(i)}$ is a
function of all the $\tau^{(a)}_j$s, and includes all color
and velocity dependence associated with the gluons, including
the vectors $\xi^\mu(\tau^{(a)}_j)$ that are contracted with
the gluon propagators.
Notice that $\F_E$ depends only on the assignment of
gluon connections, $E$, and is otherwise independent
of the internal function $\W_E^{(i)}$, including its order, $i$.   We 
now use this property of the $\F$'s to derive an identity
that will serve as a lemma for our main result.

\begin{figure}
{\hskip .5 in \epsfxsize= 15 cm \epsffile{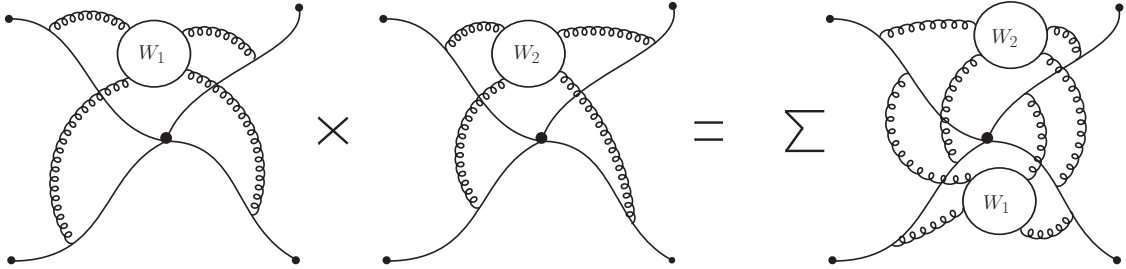}}\\
 \caption{Illustration of the coordinate identity, Eq.\ (\ref{eq:coordidentity}), where each
 line attaching $W_1$ and $W_2$ to a Wilson line stands for an
 arbitrary number of gluons ($e^{(a)}_i$ in the text.) The sum on
 the right represents the sum over all mutual orderings 
 of the external gluons of $W_1$ and $W_2$, preserving the orderings internal to each
 $W$ along the Wilson line.   The color-dependent product
 of the web internal factors, $W_1$ and $W_2$ are the same on both sides of the figure.}  
 \label{fig:identity}
\end{figure}

Let us consider the product of functionals, $\F_{E_s}$, $s=1\dots m$,
with each factor defined by (\ref{eq:calFcalW}).  
For a given choice of Wilson line $a$, the integrals within each factor of the product are ordered
as in (\ref{eq:calFcalW}) above, but they are not otherwise mutually ordered
between different products.   We can, however, write the product
as a sum of terms, in which all the integration parameters $\tau^{(a)}_{j_s}$ 
from every factor $\F_{E_s}$,  $s=1\dots m$ are ordered 
with respect to the integrals along every line from every other factor, while maintaining
the original ordering within each factor.   The sum is effectively over all possible interleaving 
of the integrals with each other.     
 We label each such ordering by $E_\pi(\cup_sE_s)$, with $\pi$ an
element of the set $\Pi(\{E_s\})$ of the permutations  
of all the parameters $\tau^{(a)}_{j_s}$, 
which preserve the original ordering internal to each $\F_{E_s}$
\bea
\prod_{s=1}^m \F_{E_s}
=
\sum_{\pi \in \Pi(\{E_s\})} \F_{E_\pi(\cup_{s=1}^m E_s)}\, ,
\label{eq:coordidentity}
\eea
This identity holds for any
sets of Wilson lines, which need not be straight, or of infinite length.
We note that at this stage, every term on the right-hand side of (\ref{eq:coordidentity})
is different, because the integrals within each $\F_{E_s}$ 
will act on different functions.  We will come back to this point shortly.
Figure \ref{fig:identity} illustrates Eq.\ (\ref{eq:coordidentity}), where the sum in
the figure represents the sum over all interleavings of gluons connecting the two $w$'s to the lines.
As the figure shows, on the right-hand side of Eq.\ (\ref{eq:coordidentity}) we have
a set of terms whose integrals can be identified with those of $N+1$st order 
diagrams.   Because they act on color-dependent web factors, however, which
do not correspond to the resulting diagrams in general, we will end up
combining diagrams with nonstandard color factors.

It may also be helpful to write a simple example of Eq.\ (\ref{eq:coordidentity}), for one line,
involving a single integral with two mutually ordered integrals, written in the notation 
introduced above,
\bea
\int_0^\infty d\tau^{(a)}_{j_1}\times \int_0^\infty d\tau^{(a)}_{k_1}\, \int_{\tau^{(a)}_{k_1}}^\infty d\tau^{(a)}_{k_2}
&=&
\int_0^\infty d\tau^{(a)}_{j_1} \int_{\tau^{(a)}_{j,1}}^\infty d\tau^{(a)}_{k_1} \int_{\tau^{(a)}_{k_1}}^\infty d\tau^{(a)}_{k_2}
\nonumber\\
&\ & \quad + \int_0^\infty d\tau^{(a)}_{k_1}\, \int_{\tau^{(a)}_{k_1}}^\infty d\tau^{(a)}_{j_1}\, \int_{\tau^{(a)}_{j_1}}^\infty d\tau^{(a)}_{k_2}
\nonumber\\
&\ & \quad \quad +
\int_0^\infty d\tau^{(a)}_{k_1}\, \int_{\tau^{(a)}_{k,1}}^\infty d\tau^{(a)}_{k_2}\, \int_{\tau^{(a)}_{k_1}}^\infty d\tau^{(a)}_{j_1}\, .
\eea
The relation (\ref{eq:coordidentity}) is simply a generalization of this trivial rewriting.

We are now ready to go back to the expression, Eq.\ (\ref{eq:first_wN+1}) for $w^{(N+1)}$.  
We expand each factor, $w^{(i)}$ as a sum over $E$, to which
we apply the representation (\ref{eq:calFcalW}),   
\bea
w^{(N+1)} 
&=& 
\sum_{D^{(N+1)}}  D^{(N+1)} \ -\
\nonumber\\
&\ & \hspace{-10mm}  \left[\, \sum_{m=2}^{N+1} \frac{1}{m!}\, 
 \prod_{j=1}^m\, \left(\, \sum_{i_j=1}^N\sum_{E_j\in E[w^{(i_j)}]} \,\right)
 \,  
\F_{E_m} [\W_{E_m}^{(i_m)}] \dots \F_{E_1} [\W_{E_1}^{(i_1)}] \right]^{(N+1)}\, .
\label{eq:second_wN+1}
\eea
This enables us to use the integration identity
(\ref{eq:coordidentity}) to combine the $\F$s, which now act on the
product of the $\W$s,
\bea
w^{(N+1)} 
&=& 
\sum_{D^{(N+1)}}  D^{(N+1)} \ -\
\nonumber\\
&\ & \hspace{-20mm}  \left[\, \sum_{m=2}^{N+1} \frac{1}{m!}\,
 \prod_{j=1}^m\, \left(\, \sum_{i_j=1}^N\sum_{E_j\in E[w^{(i_j)}]} \,\right)
 \, 
\sum_{\pi \in \Pi(\{E_s\})} \F_{E_\pi(\cup_{s=1}^m E_s)}\, [\W_{E_m}^{(i_m)} \dots \W_{E_1}^{(i_1)}] \right]^{(N+1)}\, .
\label{eq:third_wN+1}
\eea
The combination of sums in this expression over the orders $i_j$ and over choices of connections of gluons, $E_j$ to the Wilson lines for each of the  $\W$s
is equivalent to the sum over all diagrams that can be formed by combing 
the $\W$s.  As above, we let $m_c$ be the number of identical factors $\W^{(i_c)}$
in Eq.\ (\ref{eq:second_wN+1}).   
The $m_c!$ sets of diagrams found by permuting the roles of the
integrals and internal factors $\W^{(i)}$ in these classes of identical $\W$s are also identical.
On the other hand, sets of diagrams found by permuting distinguishable internal factors are also distinguishable.
We may therefore replace the product of sums over orders and gluon connections
of the internal factors in
Eq.\ (\ref{eq:third_wN+1}) with a sum over all distinguishable 
permutations of the integration factors in $\Pi(\{E_s\})$ found
by combining the $\F$s, taking into account the distinguishability properties of the $\W$s.
We need only weight this sum by the factor $\prod_c m_c!$.

Once the sum over permutations $\pi$ in (\ref{eq:third_wN+1}) is over distinguishable 
combinations of the integrals and $\W$s, each choice of $\pi$ uniquely determines 
a set of diagrams, found by summing over each of the internal factors $\W_{E_j}^{(i_j)}$
while leaving the orderings of its connections to the Wilson lines fixed.  
We can therefore
replace the sum over $\pi$ with a sum over distinguishable $N+1$st-order diagrams
in (\ref{eq:third_wN+1}), found from the lower-order
internal functions $\W_E^{(i)}$, combined with a sum over all possible ways of forming that diagram
from a product of internal functions $\W$ of lower order.  

To be specific, we write
$D^{(N+1)} = \sum_E D_E^{(N+1)}$, where the sum is over all possible connections
$E$ to the Wilson lines for $N+1$st order diagrams $D_E^{(N+1)}$.
Let $\Omega_m(D_E^{(N+1)})$
be the set of all combinations of $m$ $\W$s that give diagrams
that are topologically equivalent to $D_E^{(N+1)}$ in this manner.
That is, only those combinations of $E_j$ and $i_j$ that reproduce the integrals of $D_E^{(N+1)}$
are included in the sum  over $\Omega_m(D_E^{(N+1)})$.
Their color factors, of course, remain defined by the original product.
Then Eq.\ (\ref{eq:third_wN+1}) may be rewritten as
\bea
w^{(N+1)} 
&=& 
\sum_{E}
\sum_{D_E^{(N+1)}}  \Bigg( \, D_E^{(N+1)} 
 - \F_E  \left[\, \sum_{m=2}^{N+1}\,
\sum_{\Omega_m(D_E^{(N+1)})} 
 \,  \frac{\prod_c m_c!}{m!}\, \sum_{sym}\, \W_{E_m}^{(i_m)} \dots \W_{E_1}^{(i_1)}\right ] \
\Bigg)\, ,
\nonumber\\
\label{eq:fourth_wN+1}
\eea
where $ \F_E$ represents the integrals along the Wilson lines for diagram $D_E^{(N+1)}$,
and $\sum_{sym}$ indicates the sum over all permutations of
the factors $\W_{E_j}^{(i_j)}$ in set $\Omega_m(E)$.
The symmetric sum is over distinguishable
permutations only, so that if two $\W$s are identical, there 
is only a single term.   Equation (\ref{eq:fourth_wN+1}) is our basic
recursive  result: from each diagram at order $N+1$
we subtract a specific set of diagrams with the same integrals over Wilson lines,
but times a color-symmetrized product of lower order internal web factors.

\begin{figure}
{\hbox{\hskip 4.5 cm \epsfxsize=4 cm \epsffile{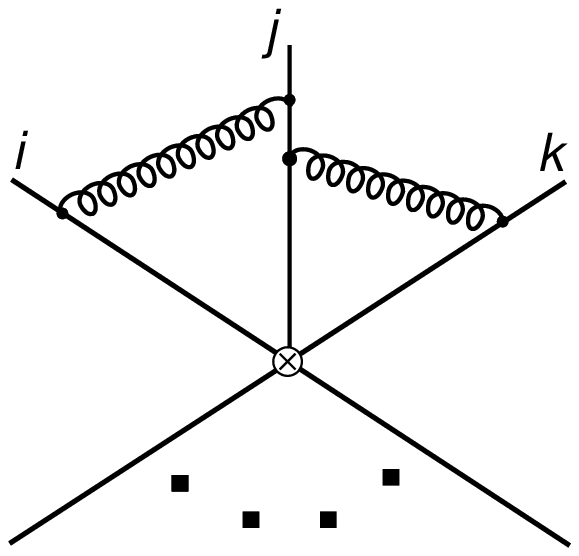} \quad \quad
\epsfxsize=4 cm \epsffile{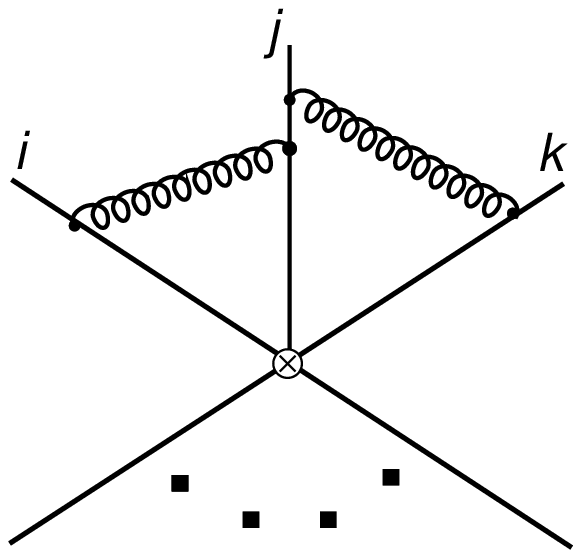}} 
\hbox{ \hskip 6 cm (a) \hskip 3.9  cm \quad (b) } \caption{Double
exchange diagrams discussed in connection with Eq.\ (\ref{eq:w2w3}).   The
shaded circle represents a multi-eikonal  vertex.}  
\label{fig:cancelpair}}
\end{figure}

The result we have just derived generalizes the web construction to
arbitrary products of Wilson lines.
It is apparently more complex than the original construction
for the cusp vertex, because of the non-commutativity
of the lower order webs.   For the cusp vertex \cite{Gatheral:1983cz}, 
the simple condition on web diagrams is that
they be irreducible under cuts of two Wilson lines.   This criterion does not 
extend to the general case.   In particular,  the diagrams in Fig.\ \ref{fig:cancelpair}, whose color factors
do not commute survive in two-loop generalized webs. 

If the color factors commute in Eq.\ (\ref{eq:fourth_wN+1}), however,
we recover the familiar web formulas.   
This is the case for an abelian theory
or for a special case in a nonabelian theory,
such as Wilson lines coupled at (singlet)  two- or three-eikonal 
vertices.  
More generally, there is a cancellation for all diagrams 
that can be decomposed into commuting subdiagrams.
Let us see how this comes about.

\begin{figure}
{\hskip 4 cm  \epsfxsize= 4.5 cm \epsffile{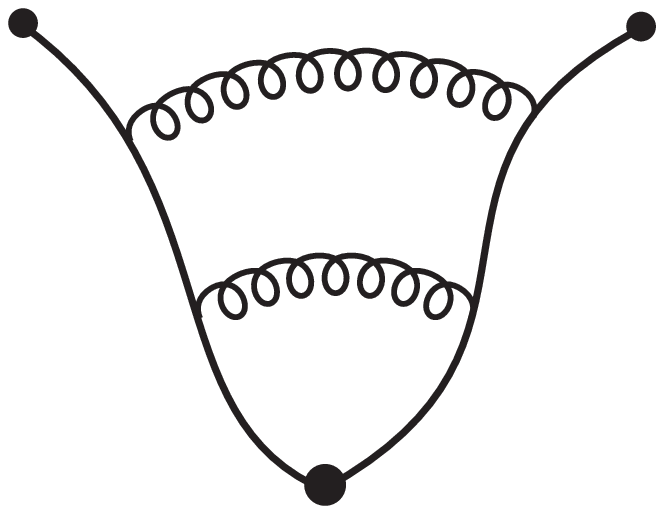} \quad \quad  \epsfxsize= 4.5 cm \epsffile{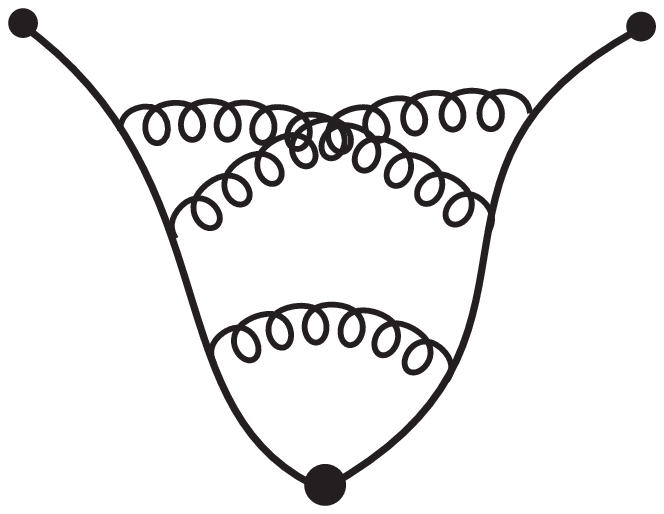}}\\
\hbox{\hskip 6 cm (a) \hskip 4.7 cm (b)}
 \caption{Examples of two-web diagrams for the form factor.}  
 \label{fig:twoeikonal}
\end{figure}

For commuting internal functions $\W$
in (\ref{eq:fourth_wN+1}), we can combine all terms in
the symmetric sum over color structures that give the same result,
remembering that terms related by permutations of identical factors $\W$ 
are counted only once.  We then find 
\bea
w^{(N+1)}_{commuting} 
=
\sum_E \sum_{D_E^{(N+1)}}  \Bigg( \, D_E^{(N+1)} \ -
  \F_E\, 
   \left[\, \sum_{m=2}^{N+1}\,
\sum_{\Omega_m(D^{(N+1)})} 
 \,  \prod_j\,  \W_{E_j}^{(i_j)} \right ] \ \Bigg)\, .
 \label{eq:commuting}
\eea
This result is formally equivalent to the usual web formula
for the form factor and related cross sections \cite{Gatheral:1983cz,Laenen:2008gt}.
We should check, however, that 
the expression vanishes for diagrams $D^{(N+1)}$
in Eq.\ (\ref{eq:commuting}) that do not appear in
the web formulation.   In the case of the cusp vertex,
webs are defined by irreducibility under cuts of the two Wilson 
lines.   Two examples are shown in Fig.\ \ref{fig:twoeikonal}.
Neither Fig.\ \ref{fig:twoeikonal}a nor b is a web, because in both cases we can cut the diagram between
the two exchanged gluons in a and between the inner gluon and the crossed ladder in b.
We will use these diagrams to sketch a demonstration of the result at all orders.

For the two Wilson line case, we
assume that up to $N$ loops all webs
are two Wilson line irreducible, and  we consider
what happens when $D^{(N+1)}$ is not such a diagram.
For definiteness, let $D^{(N+1)}$ consist of a ladder with two
rungs, as in Fig.\ \ref{fig:twoeikonal}.  

Suppose first that the two rungs
have no web subdiagrams.   In this case the color factor
for diagram $D^{(N+1)}$ is the product of
the (two) web color factors, and
is clearly cancelled by the product of web color factors
in Eq.\ (\ref{eq:commuting}).   This is the case for Fig.\ \ref{fig:twoeikonal}a.

Next suppose that the rungs themselves 
are decomposable into sets of webs.   An example 
is Fig.\ \ref{fig:twoeikonal}b, because the outer, crossed ladder,
although a web, is formed
of two single-gluon exchanges, each of which can play the role of a first-order web.
In this case, the color factors of the $\W$s are
not the same as in the original diagram $D^{(N+1)}$,
but have additional subtractions, corresponding
to each decomposition into webs.
For each such decomposition, however, there is also a corresponding term
in the original sum over web decompositions
of $D^{(N+1)}$.   For our example, this is the unique three-web decomposition
of Fig.\ \ref{fig:twoeikonal}b into three webs.   Thus, here
also, this diagram does not contribute to $w^{(N+1)}$.   
This pattern clearly continues for
more and more complex ladders for the form factor.

\section{Renormalization of Multi-eikonal Vertices}
\label{sec:ren}

In the following we clarify the systematics of the all-order renormalization of multi-eikonal vertices. 
As we demonstrate, in the presence of non-commuting color tensors, renormalization becomes more 
complicated, yet remains tractable. Our results generalize to all orders the non-trivial two-loop 
contribution to the anomalous dimension matrix for massive partons found recently 
\cite{Ferroglia:2009ep,Mitov:2010xw} with the help of a direct calculation. 
We recall that at two loops in the purely massless case no such non-trivial contributions 
involving counterterms appear \cite{Aybat:2006mz,Mitov:2010xw}. Clearly, it will be very interesting 
to understand if the apparent simplicity observed in the massless case 
both in QCD and related theories is accidental or, if found to persist through higher orders, is due to some deeper reason.

Our starting point is the all-order multiplicative renormalization property of the effective multi-eikonal vertex 
\bea
A_\ren\ =\ A\, Z^{-1}\ \equiv\ \exp [w]\, \exp [\zeta]\, .
\label{eq:Aren}
\eea
Note that the ``un-renormalized'' vertex $A$ contains UV renormalization 
for sub-divergences not related to the effective vertex (i.e. coupling renormalization).\footnote{We note that for eikonal Wilson lines, beyond zeroth order $A$ is defined by scaleless integrals, and formally vanishes in dimensional regularization.  This is not the case, however, for more general Wilson loops.}
Moreover, we observe that  $w$ and $\zeta$ are, in general, color matrices and 
do not commute with each other. 
Combining the two exponentials appearing in Eq.~(\ref{eq:Aren}), and introducing the perturbative expansions
\bea
w = \sum_{i\ge 1} w^{(i)}\, , \quad \zeta = \sum_{i\ge 1} \zeta^{(i)}\, ,
\eea
we can define all-order ``renormalized webs" by 
\beq
A_\ren = \exp\Bigg\{ w + \zeta + H(w,\zeta)\Bigg\}
= \exp\Bigg\{ \sum_{i\ge 1} \left[w^{(i)} +  \zeta^{(i)}\right] +  
\sum_{j\ge 2} H^{(j)}(w,\zeta)\Bigg\}\, ,
\label{eq:Aren-combined-exp}
\eeq
where the matrix $H(w,\zeta)$ and its perturbative expansion $H^{(j)}(w,\zeta)$ 
follow from the usual Baker-Campbell-Hausdorff  series,
\bea
H(w,\zeta) =
\frac{1}{2}\, \left[ w,\zeta\right] + \frac{1}{12}\left[ w,\left[w,\zeta\right]\right] - 
\frac{1}{12}\left[ \zeta,\left[w,\zeta\right]\right]
+ \dots
\eea
Therefore, for any theory, the amplitude can be written as the exponential 
of ``renormalized" web functions, defined at $n$th order by
\bea
w^{(n)}_\ren =  w^{(n)} +  \zeta^{(n)} + H^{(n)}(w,\zeta)\, .
\eea

The commutators $H$ are absent in any amplitude where the webs commute in their color content. 
In gauge theories these are the singlet products of one and two webs (singlet form factors). 
Therefore, in such special cases, the terms $\zeta^{(n)}$ act as local counterterms directly. This is the case
for the cusp vertex, and in this and similar cases, renormalization proceeds by simple
addition in the exponent. 
Strikingly, a similar feature holds in large-$N$ QCD, for which the 
commutator terms are non-leading in the number of colors $N$.
In this case, a multi-eikonal vertex breaks up into sums of cusps, 
and the complete web is the sum of all cusp exponents,
each of which is renormalized additively.

Turning to the general case, we need to clarify how the $\zeta$'s determine the anomalous dimensions. 
For any function $f(\as(\mu))$, we define the derivative with respect to the scale $\mu$ as
\beq
f'(\as)  \equiv \mu\frac{d}{d\mu}\, f(\as(\mu)) = \left[-2\vep \alpha_s  + \beta(\as)\right] \,  \frac{\partial}{\partial\as}\, f(\as(\mu)) 
\label{eq:derivative}
\, ,
\eeq
with the second form for a dimensionally-regulated theory in $D=4-2\vep$ dimensions, and $\beta(\as)=-\as^2/(2\pi)\sum_{n=0}^\infty \beta_n(\as/\pi)^n$ with $\beta_0=11C_A/3-2n_f/3$.

The anomalous dimensions  can be
found from the relation (see also Ref.~\cite{Aybat:2006mz})
\beq
\mu\frac{d}{d\mu}\, A_{\rm ren} =  -\ A_{\rm ren}\, \Gamma =   e^{ w } \, \mu\frac{d}{d\mu}\, e^{ \zeta} 
=  e^{ w }\, e^{ \zeta}\, \sum_{k=0}^\infty \frac{(-1)^k}{(k+1)!}\, {\cal C}_k(\zeta,\zeta') \, , 
\eeq
with $\Gamma=Z^{-1}\mu dZ/d\mu$ and with the matrices ${\cal C}_k$ defined recursively
as nested commutators, 
\bea
{\cal C}_0(\zeta,\zeta') &=& \zeta'\, ,
\nonumber\\
{\cal C}_{k+1} &=& [\zeta,{\cal C}_k(\zeta,\zeta')] \, .
\eea
Thus, to any order, the anomalous dimension matrix is found from the 
single pole of the $n$th-order counterterm, plus nested commutators of lower order counterterms,
\bea
\Gamma^{(n)}  =  - (\zeta')^{(n)} - \left( \sum_{k=1}^\infty \frac{(-1)^k}{(k+1)!}\, {\cal C}_k(\zeta,\zeta')\, \right)^{(n)} \, .
\label{eq:Gamma-n}
\eea
For the cusp, and related cases where there is no mixing between color structures, the
nested commutators vanish, and the anomalous dimension is determined directly from $\zeta'={\cal C}_0$.

It is instructive to see low-order examples. At two and three loops we have
\bea
w_\ren^{(2)} &=& w^{(2)} + \zeta^{(2)} + \frac{1}{2}\, \left[\, w^{(1)},\zeta^{(1)}\, \right] \, , \nonumber\\
w_\ren^{(3)} &=& w^{(3)} + \zeta^{(3)} + \frac{1}{2}\, \left(\, \left[\, w^{(1)},\zeta^{(2)}\, \right] 
+ \left[\, w^{(2)},\zeta^{(1)}\, \right] \, \right)
\nonumber\\
&\ & \quad +\ \frac{1}{12}\,  \left[\, w^{(1)}-\zeta^{(1)},\left[w^{(1)},\zeta^{(1)}\right]\right] \, ,
\label{eq:w2w3}
\eea
corresponding, with the help of Eq.~(\ref{eq:Gamma-n}), to anomalous dimensions 
\bea
\Gamma^{(2)}  &=& - (\zeta')^{(2)} \, , 
\nonumber\\
\Gamma^{(3)}  &=& -(\zeta')^{(3)} + \frac{1}{2} \left[\zeta^{(1)},(\zeta')^{(2)}\right]
+ \frac{1}{2} \left[\zeta^{(2)},(\zeta')^{(1)}\right]\, .
\label{eq:Gammazeta}
\eea
The structure of the two-loop web $w_\ren^{(2)}$ is very transparent in light of the results of Ref.~\cite{Mitov:2010xw}. 
In terms of the notation given here, it was found that (in Feynman gauge) $w^{(2)}$ gets no contributions from the double exchange diagrams illustrated by Fig.\ \ref{fig:cancelpair}, but that the full two-loop counterterm, $\zeta^{(2)}$, and therefore $\Gamma^{(2)}$ \cite{Ferroglia:2009ep} requires diagrams in which the inner loops of Fig.\ \ref{fig:cancelpair} are replaced by one-loop counterterms.  In Eqs.\ (\ref{eq:w2w3}) and (\ref{eq:Gammazeta}), these terms are generated by
 the commutator between the derivative of the one-loop counterterm $\zeta^{(1)}$ and the one-loop web $w^{(1)}$.   Because the poles of the one loop web are proportional to the poles of $\zeta^{(1)}$, there are no double poles in the commutator. 
We postpone the detailed analysis of the three-loop web $w_\ren^{(3)}$ for future work.

In applying the procedure sketched above at fixed order, the  ``practical"
approach in deriving renormalized webs, counterterms and anomalous dimension matrices is, first to expand the all-order exponent Eq.~(\ref{eq:Aren-combined-exp}) and to compare the $n$th order of this expansion to the usual diagrammatic prescription for the calculation of the corresponding amplitude at the same order. This is analogous to the use of Eq.\ (\ref{eq:first_wN+1}) for unrenormalized webs.  Then, requiring that $w_{\ren}^{(n)}$ be finite fixes the counterterm $\zeta^{(n)}$ in terms of the webs of the same order and a combination of webs and counterterms of lower orders. Finally, one can determine the anomalous dimension (matrix) $\Gamma^{(n)}$ in terms of the $n$th order counterterm $\zeta^{(n)}$ through Eq.~(\ref{eq:Gamma-n}).

\section{Extensions and Conclusions}
\label{sec:extend}

In this brief account, we have shown that a large class of products of Wilson lines and loops share a straightforward diagrammatic construction that results in an exponentiated form.  This construction of an exponent generalizes the webs of color-singlet form factors and related cross sections \cite{Gatheral:1983cz}.   For general multi-eikonal vertices, the diagrams necessary to compute the exponent do not share quite the simple rule of irreducibility under cuts of Wilson lines.   In fact, this feature was already clear on the basis of two-loop renormalization for vertices connecting several massive eikonal lines \cite{Ferroglia:2009ep,Mitov:2010xw}.

The construction is based on counting, and is carried out in coordinate space.    The underlying identity, Eq.\ (\ref{eq:coordidentity}), between products of integrals along the paths of ordered exponentials, reduces to the Fourier transform of the momentum space eikonal identity \cite{Gatheral:1983cz} in the limit of straight, semi-infinite paths, but is much more general.  It applies as well to closed paths, with or without sequential cusp singularities, so that the diagrammatic construction described here applies in those cases as well.

We have seen in Sec.\ \ref{sec:ren} that diagrammatic exponentiation is most direct before the renormalization of the multi-eikonal vertices.  The renormalization of such a vertex leads to a fairly complex, but highly structured, formalism for determining anomalous dimensions.    We anticipate that the investigation of generalized webs at higher orders will shed further light on matrices of anomalous dimensions that appear in many resummed cross sections, and perhaps on the dynamics of soft interactions in gauge theory more generally.

The extension of these rules to eikonal cross sections involving the scattering of Wilson lines with arbitrary colors is possible, because in squared amplitudes involving products of ordered exponentials, we can follow the path of a line that extends from a multi-eikonal vertex into the final state to infinite time in the amplitude, and then back again to the vertex in the complex conjugate amplitude.    Because the arguments given above apply to any path, and can accommodate any form of propagators and interactions between gluons, they extend to QCD hard scattering in eikonal approximation, just as the web formalism extends to two-jet cross sections, to DIS, and to Drell-Yan annihilation processes.

In conclusion, we touch on the zero-mass limit for the generalized web construction.   For soft functions in resummed cross sections, the soft function is conveniently described as a ratio of eikonal amplitudes or cross sections divided by eikonal form factors \cite{Sterman:2002qn} or jet functions \cite{Berger:2003iw}, respectively.   In such ratios, double poles and logarithms associated with collinear behavior cancel.   Because the form factors and jet functions exponentiate according to the original web construction, the complexities associated with color structure remain in the soft function, which exponentiates single logarithms in matrix form.    If the conjecture that the anomalous dimensions of massless Wilson lines reduce to a dipole structure only holds, in this limit the general construction will simplify to a sum of exponentiated webs \cite{Becher:2009qa,Gardi:2009qi}.   The general pattern of higher order counterterms described here may help in the investigation of this possibility.

\noindent
{\bf Note added:} During the completion of this project, we learned that E.\ Gardi, E.\ Laenen, G.\ Stavenga and C.\ White were completing a related study on the generalization of webs~\cite{Gardi:2010rn}.

{\bf Acknowledgments:} 
This work was
supported in part by the National Science Foundation, grants
PHY-0354776, PHY-0354822 and PHY-0653342. The work of A.M. was supported by a
fellowship from the {\it US LHC Theory Initiative} through NSF grant
0705682.  We thank Lance Dixon, Einan Gardi, Eric Laenen and Lorenzo Magnea
for very interesting conversations, and for sharing some of their unpublished work.

\end{document}